      [1999/12/01 v1.4c Il Nuovo Cimento]
\documentclass{cimento}

\def\lesssim{\mathrel{\hbox{\rlap{\hbox{\lower2pt\hbox{$\sim$}}}\raise2pt\hbox{$<$}}}}
\def\grtsim{\mathrel{\hbox{\rlap{\hbox{\lower2pt\hbox{$\sim$}}}\raise2pt\hbox{$>$}}}}
\newcommand{\gtsim}{\mbox{{\raisebox{-0.4ex}{$\stackrel{>}{{\scriptstyle\sim}}$}}}}
\newcommand{\ltsim}{\mbox{{\raisebox{-0.4ex}{$\stackrel{<}{{\scriptstyle\sim}}$}}}}

\usepackage{graphicx}  % got figures? uncomment this
\title{Accretion indicators for the 37 brightest radio sources in the Subaru/\textit{XMM-Newton} Deep Field}
\author{E.~Vardoulaki\from{ins:1}\ETC,
S.~Rawlings\from{ins:1} 
\atque 
C.~Simpson\from{ins:2}}
\instlist{\inst{ins:1} Astrophysics, Denys Wilkinson 
Building, Keble Road, Oxford, OX1 3RH, UK.
\inst{ins:2} Astrophysics Research Institute, Liverpool John Moores
University, Twelve Quays House, Egerton Wharf, Birkenhead CH41 1LD, UK.
}
\begin{document}

\maketitle

\begin{abstract}
We study the 37 brightest radio sources in the Subaru/\textit{XMM-Newton} 
Deep Field (SXDF). Using mid-IR (Spitzer MIPS 24 $\mu \rm m$) data we expect 
to trace nuclear accretion activity, even if it is obscured at optical 
wavelengths, unless the obscuring column is extreme. Our results suggest that 
above the `FRI/FRII' radio luminosity break most of the radio sources are 
associated with objects that have excess mid-IR emission, only some of which 
are broad-line objects, although there is one clear low-accretion-rate FRI. 
The fraction of objects with mid-IR excess drops dramatically below the 
FRI/FRII break, although there exists at least one high-accretion-rate QSO. 
Investigation of mid-IR and blue excesses shows that they are correlated as 
predicted by a model in which a torus of dust absorbs $\sim$30\% of the light, 
and the dust above and below the torus scatters $\gtsim$1\% of the light.
\end{abstract}

\section{Introduction and Data}

Powerful radio sources are believed to have central super-massive black holes 
(SMBH) with uniformly high accretion rates at the highest radio luminosities 
and typically lower accretion rates at lower radio luminosities (\cite{rs91}). 
Low-luminosity radio jets can, however, be associated with high-accretion-rate 
systems, and these so-called `radio quiet' quasars appear to have similar 
FRI-like radio structures to low-accretion-rate counterparts of similar radio 
luminosity (e.g. \cite{hbr07}). At low redshift, the most massive 
($\gtsim$ $10^{8} \rm M_{\odot}$) SMBH typically have very low accretion rates 
with systematically higher average values at $z \gtsim$ 2, the so-called 
`quasar epoch' (\cite{yt02}). These observational results fit in with 
theoretical ideas that a `quasar mode' of feedback is prevalent in the distant 
universe, and that a `radio mode' feedback is dominant at low redshift 
(e.g. \cite{cro06}). 

The central region of an AGN is surrounded by a dusty torus which absorbs 
light and re-emits it in the infrared. Above and below the plane of the torus, 
dust scatters light yielding a blue excess. Such mechanisms make it difficult 
to observe objects viewed through the torus directly in the optical, UV and 
soft X-rays. The torus creates anisotropic obscuration of the central regions 
resulting in two different types of observed objects, type 1 that are viewed 
face-on and type 2 that are viewed edge-on. Here we use mid-IR observations 
to search for evidence of accretion in a manner which is far less dependent 
on orientation.

The sample studied here is the 37 brightest radio sources from the VLA survey 
of the Subaru/\textit{XMM-Newton} Deep Field (SXDF; \cite{sim06}) with 
flux densities greater than 2 mJy at 1.4 GHz. Optical, X-ray and radio 
observations of the SXDF were made within the 1.3 square degree 
Subaru/\textit{XMM-Newton} Deep Field with Subaru, \textit{XMM-Newton} and the 
VLA respectively. Thirteen of our objects are not as yet spectroscopically 
confirmed so we use photometric redshifts in these cases, calculated with the 
HYPERz code (\cite{boz00}) and typically nine data points from $B-$band 
(440 nm) to 4.5 $\mu \rm m$ (Vardoulaki et al. in prep).

\section{Discussion}

\begin{figure}
\begin{center}
\setlength{\unitlength}{1mm}
\begin{picture}(140,38)
\put(65,-21){\includegraphics{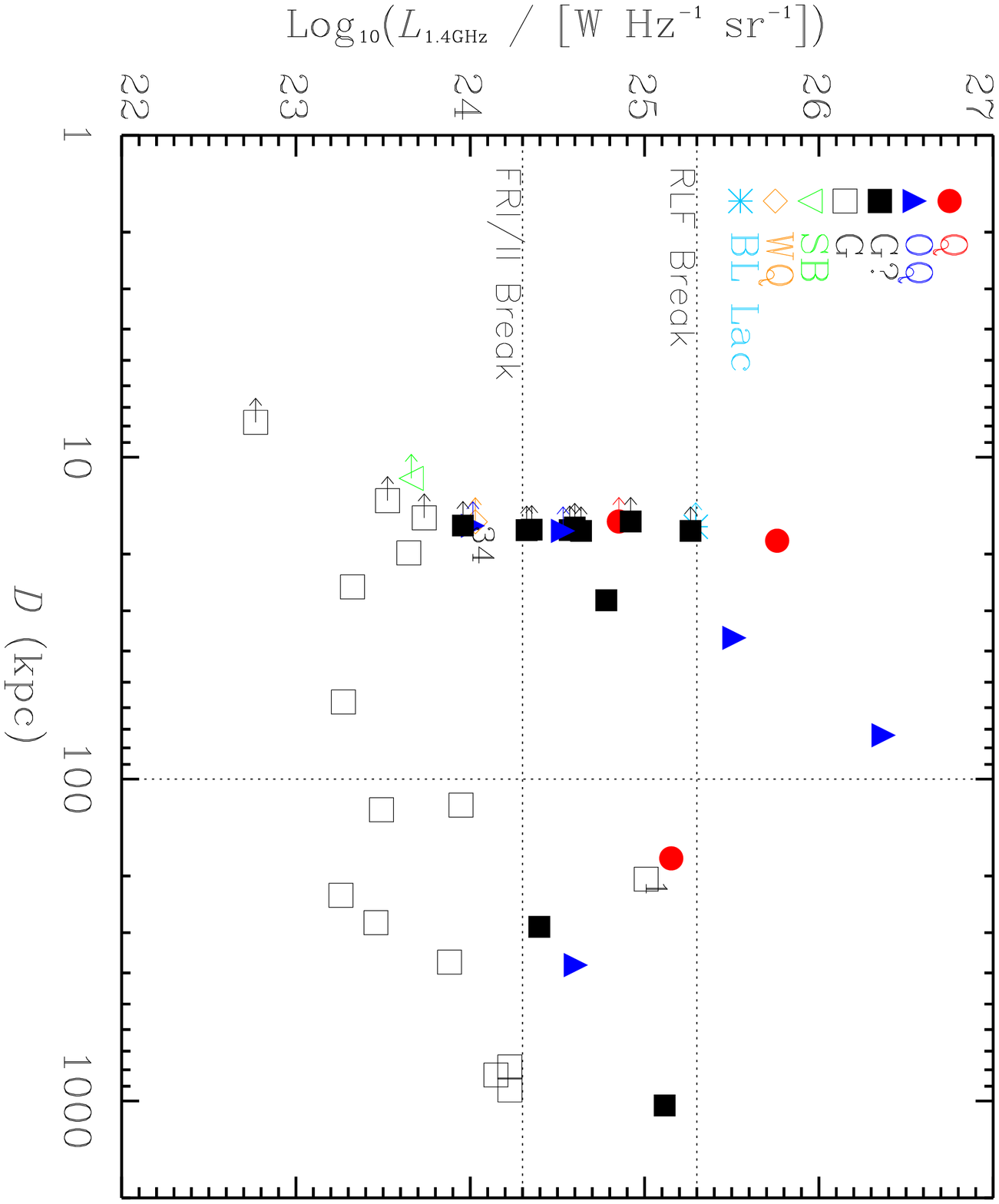}}
\put(143,-21){\includegraphics{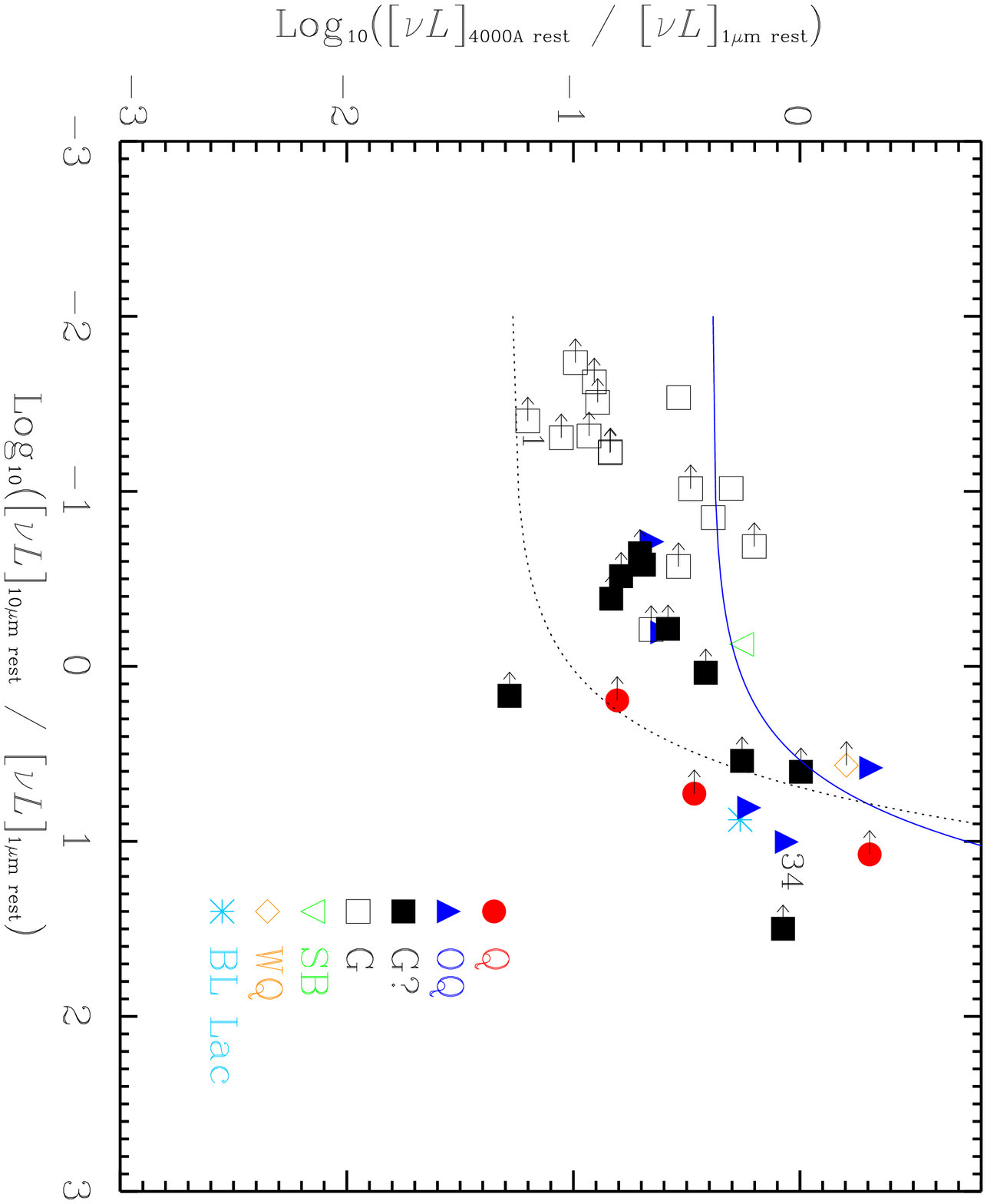}}

\end{picture}
\end{center}
\vspace{0.5in}
{\caption[junk]{\label{fig1}
a {\it Left}: Radio Luminosity 
$\log_{10}(L_{1.4{\rm GHz}}/ \rm [W Hz^{-1} sr^{-1}])$ versus 
largest projected linear size $D$: symbols indicate optical/IR 
classification; filled red circles for quasars `Q'; filled blue triangles for 
obscured quasars `OQ'; filled black squares for possible galaxies `G?'; black 
squares for secure galaxies `G'; green upside-down triangles for starbursts 
'SB'; orange diamonds for weak quasars `WQ'; and light blue stars for BL Lac 
`BL'. The horizontal lines show the RLF and FRI/FRII breaks calculated from 
the values in \cite{fr74} using a typical steep-spectrum spectral index of 0.8 
and translated to our assumed cosmology. b {\it Right}: Blueness versus mid-IR 
excess. Symbols are the same as in the left figure. The black dotted line 
corresponds to the best-fit line in the log-linear plane where all objects, 
were treated as detections; the slope and intercept are 0.26 and -1.27 
respectively, giving (see Eqn (2)) $\rm k_{1} = 0.05$ and $\rm k_{2} = 0.03$. 
The blue solid line corresponds to the best-fit line in log-linear plane 
where objects without detections at 24 $\mu \rm m$ were treated as limits; the 
slope and intercept are 0.10 and -0.41 respectively, giving 
$\rm k_{1} = 0.39$ and $\rm k_{2} = 0.09$. The Buckley-James method in the 
ASURV statistics package (\cite{lav92}) was used in these calculations. `SB' 
and `BL' objects were excluded from the calculations since they have SEDs 
dominated by different physical processes to those assumed in the model 
described by Eqns (1) and (2). We adopt a radio spectral index $\alpha$ = 0.8 
($S_{\nu} \propto \nu^{-\alpha}$), unless a spectral index could be calculated 
using 1.4 GHz data from \cite{sim06} and 325 MHz data from \cite{tasse} (see 
Vardoulaki {\it et al.} in prep.). We assume throughout a low-density, 
$\Lambda$-dominated Universe in which 
$H_{0}=70~ {\rm km~s^{-1}Mpc^{-1}}$, $\Omega_{\rm M}=0.3$ and 
$\Omega_{\Lambda}=0.7$.
}}
\end{figure}

\addtocounter{figure}{0}

We use optical/IR observations to classify a radio source as either Quasar 
`Q', Obscured Quasar `OQ', Galaxy? `G?', Galaxy `G', Starburst `SB', Weak 
Quasar `WQ' or BL Lac `BL'. We deem that nuclear accretion is `significant' 
in objects that obey 
$\log_{10}(L_{24 \mu \rm m}/ \rm [W Hz^{-1} sr^{-1}]) > 23.1$ (or 
$[\lambda L]_{24 \mu \rm m} > 10^{37.3}$ ${\rm W}$). This value corresponds to 
$[\lambda L]_{24 \mu \rm m} \ge 10^{-1.8} L_{Edd}$, a typical lower limit for 
quasars (\cite{mcl04}), for a black hole mass 
$M_{\rm BH} \ge 10^{8} M_{\odot}$, a typical lower limit for radio sources 
(\cite{mcl_etal04}); $L_{Edd}$ is the Eddington luminosity. We then define the 
following categories:\\
i) {\bf Q}: Broad lines in the optical spectrum (3/37 cases). None 
of these are detected at 24 $\mu \rm m$, although their limits are 
insufficient to rule out significant accretion.\\
ii) {\bf OQ}: Objects with a 24-$\mu \rm m$ detection (5/37 cases) and with 
sufficient $L_{24}$ to represent significant accretion. This 
class may be incomplete in that some objects in the `G?' class, as 
described next, have limits above this critical value.\\
iii) {\bf G?}: A galaxy that has a 24-$\mu \rm m$ limit consistent with it 
lying above the 
$\log_{10}(L_{24 \mu \rm m}/$ $\rm [W Hz^{-1} sr^{-1}]) = 23.1$ 
line\footnote{Because of the 24 $\mu \rm m$ flux density limit, these objects 
are at high redshift, and hence, because of the 1.4-GHz flux density limit, a 
high-$L_{1.4 \rm GHz}$ sub-set of the objects lacking Spitzer 
24 $\mu \rm m$ detections.} (11/37 cases).\\
iv) {\bf G}: All other objects (15/37 cases) without significant accretion, 
unless they fall into three special categories defined by properties derived 
from spectroscopy, the SED and the optical structure: {\bf SB}: evidence from 
the SED of a starburst component (1/37 cases); {\bf WQ}: evidence from the 
SED of a quasar component but no 24 $\mu \rm m$ detection (1/37 cases); 
{\bf BL}: featureless red continuum and a point source at $K$ (1/37 cases).\\
%Objects labelled as `Q', `OQ' and `G?' include all objects with, or 
%potentially with, significant nuclear accretion, and are represented by 
%filled symbols in Figure 1.

Figure 1a shows the 1.4-GHz radio luminosity at $L_{\rm 1.4 GHz}$ versus the 
projected linear size $D$ with the symbols denoting the different optical/IR 
classes. We see that nearly all `Q', `OQ' and `G?' objects of our sample lie 
above the `FRI/FRII' luminosity 
break\footnote{Although the FRI/FRII classification scheme is on the basis of 
radio structure, there is a sharp change in radio structure at a 
characteristic radio luminosity \cite{fr74}.}, with 
the exception of the `OQ' sxdf\_0034 (the `G?' object near sxdf\_0034 lies 
very close to the boundary of significant accretion). In previous studies, the 
quasar fraction has been defined as the number of sources with quasar-like 
optical features (e.g. broad lines) and has a value of 
$\sim 0.1 \rightarrow 0.4$ over this range of $L_{1.4 \rm GHz}$ 
(e.g. \cite{wil00}). We introduce the `quasar-mode fraction' $f_{\rm QM}$ to 
describe the fraction of objects with high accretion rates to the total 
number of objects. Above the FRI/FRII break $f_{\rm QM} \sim 0.5 - 0.9$ (the 
lower value assumes the 24 $\mu \rm m$ limits are much higher than the true 
24 $\mu \rm m$ values, whereas the higher value assumes the true values lie 
just below the limits). The one clear exception in this regime is sxdf\_0001, 
which has no evidence of a QSO and a clear Twin-Jet (FRI) radio structure.

The quasar-mode fraction drops dramatically below the FRI/FRII 
break\footnote{We note that objects in our sample above the FRI/FRII break 
have median redshift $z_{\rm med} \sim 1.6$, whereas those below have 
$z_{\rm med} \sim 0.65$, so evolutionary effects may also be important.}, and 
whether or not one excludes some of the compact ($D <$ 100 kpc) sources as 
potentially part of a separate (beamed) population, then $f_{\rm QM} \ltsim$ 
0.1 because nearly all objects are galaxies `G'. The counter example here 
are sxdf\_0034, the only `OQ' below the FRI/FRII break, and potentially an 
optically-obscured example of unobscured FRI QSOs already studied in this 
radio luminosity regime (e.g. \cite{hbr07}).

Inspection of the SEDs (Vardoulaki {\it et al.} in prep) shows that some of 
our objects have an excess at 24 $\mu \rm m$ above that expected from 
extrapolation of the stellar populations. This is quantified via a measure of 
the mid-IR excess, 
$\log_{10}([\nu L]_{10 \mu \rm m rest} / [\nu L]_{1\mu \rm m rest})$. 
A comparison of mid-IR excess and blueness is presented in Fig. 1b 
where a positive correlation is obvious. The generalised Spearman correlation 
calculated using survival analysis statistical package ASURV (\cite{lav92}) is 
0.657 with a 99\% probability for a correlation.

Consider a simple model in which blueness is connected to mid-IR excess 
through the following equations\footnote{Equation (2) relies on the 
$ln(1+x) \approx x$ approximation which is only 
accurate around and below the knees of the functions plotted in Fig. 1b.}: 
\begin{equation}
[\nu L]_{4000 \rm \AA \rm rest} = 
\rm k_{1} \times [\nu L]_{1\mu \rm m rest}  + 
\rm k_{2} \times [\nu L]_{10 \mu \rm m rest}   \Rightarrow
\end{equation}
\begin{equation}
\log_{10}\left(\frac{[\nu L]_{4000 \rm \AA rest}}{[\nu L]_{1\mu \rm m rest}} \right) = 
\log_{10}(\rm e) \times \frac{\rm k_{2}}{\rm k_{1}} \times 
\left(\frac{[\nu L]_{10 \mu \rm m rest}}{[\nu L]_{1\mu \rm m rest}}\right) + 
\log_{10}(\rm k_{1}),
\end{equation}
where $\rm k_{1}$ encodes the contribution of the stellar population of a 
passively evolving galaxy formed at high redshift ($z > 5$), and $\rm k_{2}$ 
the mid-IR-excess parameter that we are looking to calculate for this sample 
of radio sources. This model assumes that light from the nucleus with 
intrinsic optical luminosity $L_{\rm opt}$ is i) absorbed by dust and 
re-emitted in the mid-IR generating luminosity $[\nu L]_{10 \mu \rm m rest}$ 
and ii) scattered, generating luminosity $[\nu L]_{4000 \rm \AA rest}$. Fig. 
1b shows best-fit lines for two scenarios: 1) all objects were treated as 
detections (black dotted line), and 2) objects are treated as upper limits 
according to their 24 $\mu \rm m$ detection (blue solid line), where in both 
cases `SB' and `BL' objects were excluded (Fig. 1b). Averaging these results 
we deduce $\rm k_{1} \sim 0.2$ and $\rm k_{2} \sim 0.05$, which agrees well 
with independent evidence. The value deduced for $\rm k_{1}$ is in line with 
the expectations of template spectra of galaxies which formed their stars at 
high redshift. Optical polarisation studies (e.g. \cite{kis01}) 
tell us that $[\nu L]_{4000 \rm \AA rest} \gtsim$ $0.01 [\nu L]_{\rm opt}$, 
which is consistent with our value of $\rm k_{2}$ given that QSO SED studies 
suggest $[\nu L]_{10 \mu \rm m rest} \sim 0.3 [\nu L]_{\rm opt}$ 
(\cite{rr95}). We conclude that whenever nuclear accretion is significant in 
our sample of radio sources, dust in the torus absorbs 30\% of the 
photons and dust above and below the torus scatters $\gtsim$1\% of the photons.

%The 37 brightest radio sources in the SXDF exhibit the following properties:\\
%{\bf $\bullet$} QSOs are confined largely, but not exclusively, to the area 
%above the `FRI/FRII' luminosity break and the quasar-mode fraction, defined 
%here as the fraction of objects with significant accretion rates to the total 
%number of objects and not as the number of sources with quasar-like optical 
%features, is $\sim 0.5 \rightarrow 0.9$. There are counter-examples like the 
%low-accretion FRI sxdf\_0001, but they are rare.\\
%{\bf $\bullet$} At $z \sim 0.65$ most of the `FRI'-regime sources have low 
%accretion rates and $f_{\rm QM} \ltsim 0.1$, but some high-accretion-rate 
%objects (e.g. sxdf\_0034) exist.\\
%{\bf $\bullet$} Mid-IR excess and blueness are correlated, where the 
%Spearman's generalised $\rho =$ 0.657 with a 99\% probability for a 
%correlation.\\
%{\bf $\bullet$} This correlation is consistent with a model in which 
%$\sim 30\%$ of the QSO light is absorbed by a dust torus and $\gtsim 1\%$ is 
%scattered by dust above and below the torus.\\

\acknowledgments

\end{document}